# MINING TECHNIQUES IN NETWORK SECURITY TO ENHANCE INTRUSION DETECTION SYSTEMS

Maher Salem and Ulrich Buehler

Network and Data Security Group,
Applied Computer Science,
University of Applied Sciences, Fulda, Germany
maher.salem@informatik.hs-fulda.de, u.buehler@informatik.hs-fulda.de

*ABSTRACT*

*In intrusion detection systems, classifiers still suffer from several drawbacks such as data dimensionality and dominance, different network feature types, and data impact on the classification. In this paper two significant enhancements are presented to solve these drawbacks. The first enhancement is an improved feature selection using sequential backward search and information gain. This, in turn, extracts valuable features that enhance positively the detection rate and reduce the false positive rate. The second enhancement is transferring nominal network features to numeric ones by exploiting the discrete random variable and the probability mass function to solve the problem of different feature types, the problem of data dominance, and data impact on the classification. The latter is combined to known normalization methods to achieve a significant hybrid normalization approach. Finally, an intensive and comparative study approves the efficiency of these enhancements and shows better performance comparing to other proposed methods.*

*KEYWORDS*

*Sequential Backward Search, Information Gain, Normalization, Probability Mass Function, Discrete Randome Variable.*

## 1. INTRODUCTION

Intrusion Detection System (IDS) got a serious attention in network security, especially in the last decade while several approaches have been proposed to enhance the performance of IDS and mitigate its drawbacks. Generally, IDS can be categorized into signature-based, anomaly-based and specification-based. Signature-based or misuse detection was the first IDS category that delivers a high degree of accuracy due to the matching techniques against known attacks. Obviously, this kind of IDS unable to detect unknown attacks [1]. Therefore, anomaly-based IDS investigated this problem and showed promising results that can detect unknown abnormal activities on the network, but it suffers from the high false alarm rate [14] [15]. Anomaly-based IDS has two learning methods, supervised learning where the IDS learns the network from its labeled data and detect anomalies based on that model, and the unsupervised learning, which is capable to handle unlabeled data by using clustering techniques. The third IDS category is specification-based IDS, which defines a system specification (model) and detects when behavior differ from expected [16]. The latter exploits finite state machine to build the model and it is mostly applied in mobile networks. Most likely, IDS exploit the network traffic to build a model that detects unseen attacks. However, the traffic of computer network consists of several features, which have various types such as nominal, numeric, or Boolean. IDS are not able to handle all features. Therefore, a significant feature selection is necessary to abstract the valuable network features that enhance the IDS performance. Moreover, features with numeric values have different scales such that a feature with huge values dominates other feature with small values. Therefore, data dominance eliminates the impact of small values, although they could support positively in the system. Thus, eliciting an optimum dataset from network traffic to enhance the accuracy and reduce the false positive rate of the classifier is a challenge that should be carefully achieved.

In this research paper two significant enhancements are proposed to treat data dimensionality and dominance and the variety of feature types. The first enhancement handles the dimensionality of features, it exploits floating search methods and information gain to infer the valuable network features. In addition, the second enhancement exploits the idea of discrete random variable and probability mass function to map the nominal features into numeric ones by treating them as sampling of discrete random variables for the respective ranges of symbols.

Thus, the above enhancements guarantee a significant and only numeric dataset from network traffic. However, network features have different scales, except the transferred nominal features, they have the same scale according to the second enhancement. Therefore, features with different scales will be normalized using decimal, statistical and minimum maximum normalization. In this paper, we introduce a hybrid approach, which consists of transferring nominal features and normalizing numeric ones, all to the same scale. Finally, several datasets are prepared based on the hybrid approach and then evaluated using several classifiers. This evaluation shows that these enhancements innervate the IDS.

In the next section we will discuss the previous works. Then the proposed enhancements will be illustrated in section 3. Section 4 explains results of the enhancements, which are based on the benchmark NSL-KDD dataset and a comparative and evaluative study as well. Finally, section 5 concludes the achievements of this paper.

## 2. RELATED WORKS

Researchers recently exploit data mining techniques in the area of IDS to enhance its performance. A fundamental phase in IDS is the feature selection, where important features are selected, so that the classifier reaches the best detection rate and the lowest false positive rate.

Yao-Hung Chan et al. [19] utilized the searching methods Sequential Backward Search (SBS) and sequential forward search with the Localized Generalization Error (L-GEM) as threshold criteria and Radial Basis Function Neural Network (RBFNN) as classifier. They showed that, the SBS method is more feasible in large datasets. On the other hand, Yang Li et al. [20] exploited Chi-Square with information gain. Chi-Square fulfills the requirements of the maximum entropy model for intrusion detection. However, the information gain method handles only discrete values, and that is not considered in their work. Therefore, a discretization approach in [21] is examined in the first enhancement. In addition, classifiers handle features with numeric values and same scale. This is achievable by normalizing these values, however, normalization have various techniques and methods. In this regard, researchers investigate the optimal and proper method in IDS area.

Oh et al. in [3] presented an unsupervised method to detect attacks using SOM. They have changed each nominal value in the features *"protocol_type, service"* by its decimal number as defined in IANA [22] protocol numbers assignment or port numbers, and then they normalized the dataset using minimum maximum normalization. Changing a string to decimal number without taking in advance the different spaces of both is affecting the normalization method and so the classifiers results. We cannot simply transfer a nominal value into decimal without any mapping function that declares the process of transformation. In addition, we have hundreds of services on the network and each has its weight by the amount of requesting it, so we cannot just change its value by its identification number. The proposed transferring method in this paper is comparied with the proposed method in [3] and results are discussed in section 4. In contrast, Cai et al. [4] proposed a unified normalization distance framework for numeric and nominal records. They mapped the nominal values to a categories domain such that each nominal value has a coordinate of 1 in its dimension in the real number space. However, they have ignored the occurrences of one nominal value in its space and how that affects its value in the real number space.

Yu Liping et al. [5] have evaluated several normalization methods in multi-attributes and they concluded that different evaluation purposes require different data normalization methods. Based on that, we have proposed here several normalization methods. In [6] Chakraborty G. and Chakraborty B. have introduced a novel normalization method by transferring the features to a higher dimension space, when all features in the same space then they have been normalized by dividing them on the longest feature vector. Unfortunately, this will not fulfill the purpose of normalizing the nominal features because they treated only numeric features.

A comparative study is presented in [7], but the nominal features have been investigated in supervised learning using decision tree and interactive dichotomizer 3. Ippoliti and Zhou in [8] have presented a modified normalization method using minimum maximum approach that adapts the input process of the Growing Hierarchical Self Organizing Map (GHSOM). They have not clarified the dynamic process for nominal features as well.

A challenging paper is proposed from Said et al. [9]. They showed an empirical comparative study between normalization methods in unsupervised learning with only Principle Component Analysis (PCA) and showed that log normalization is the best for PCA, but nominal values have not been treated on their proposal. Wang et al. [10] have proposed a study about normalizing network attributes; they have presented four normalization methods and evaluated them with four classifiers. The method frequency normalization has been presented in their work but it is shallowly declared. Comparing to the latter, we declare precisely and clearly the proposed normalization method in this work. In addition, we evaluate several normalized datasets using several classifiers.

## 3. PROPOSED ENHANCEMENTS

The proposed enhancements in this research concentrate on two main phases of IDS which are feature selection and normalization. Feature selection enhancement (or the first enhancement) is enhanced by an improved method that filters the most valuable features for IDS. On the other hand, the normalization of nominal features (or the second enhancement) solved the problem of different feature types, data dominance, and impact on classification. The latter is modified to be hybrid approach. These enhancements boost significantly the classifier performance. Figure 1 shows a general overview about the proposed enhancements

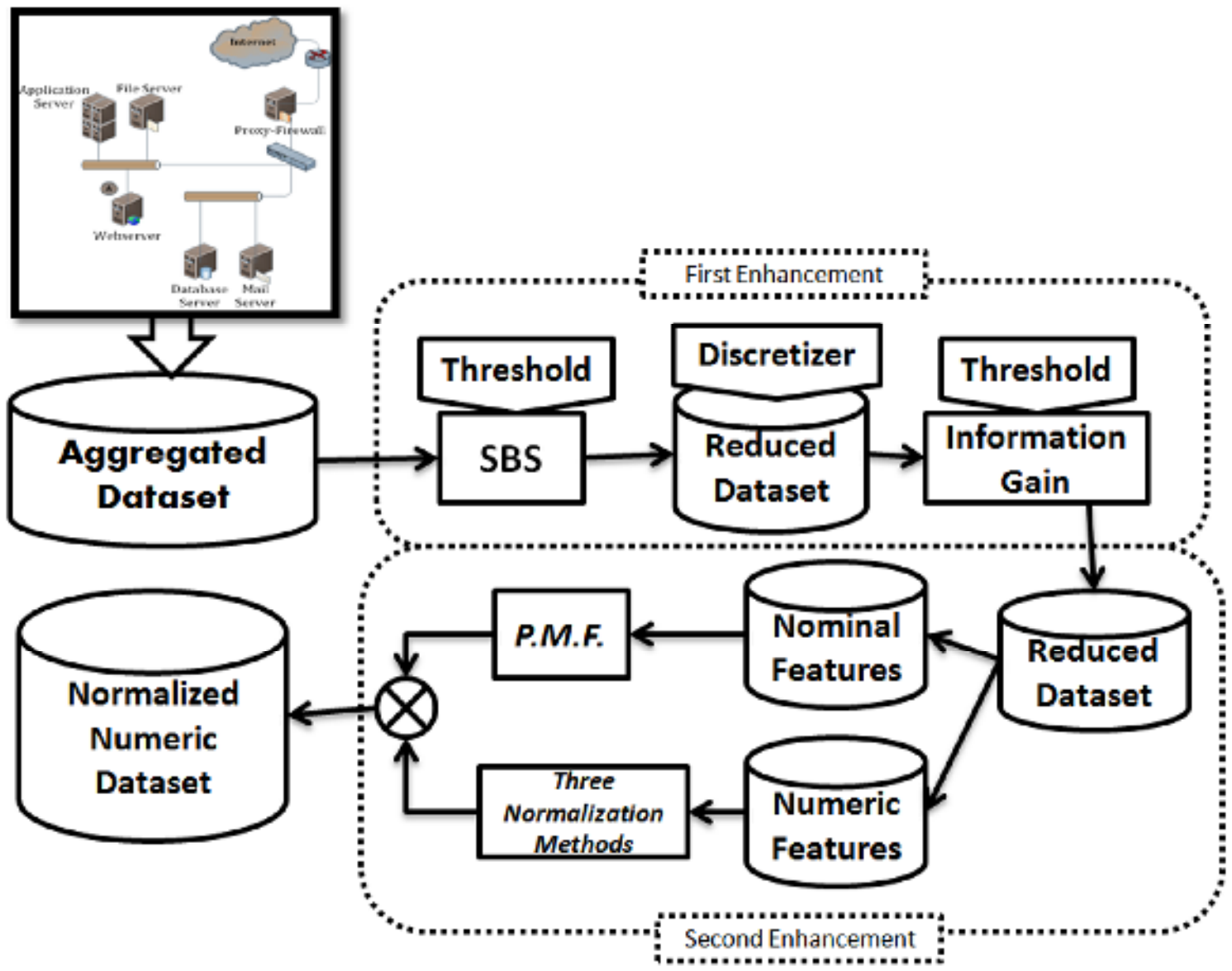

Figure 1. Overview of Proposed Enhancements

More details about these enhancements are declared in the following subsections.

## 3.1. Improved Feature Selection

To find the most valuable features, we propose the first enhancement, which is an improved feature selection method. It exploits a modified Sequential Backward Search (SBS) method and Information Gain (IG) such that only features with positive impact on classifiers and high information amount are selected. To evaluate this enhancement, we initialize a dataset with all 41 features from NSL-KDD [23] dataset and classify it using several classifiers; these are Multilayer Perceptron, Naïve Bayes, Random Tree, and Decision Tree [13]. The following steps are explaining the modified SBS method:

---

1. Initialize $\mathbb{D}$ as full data set with $\mathbb{F}$ features and $\mathbb{I}$ instances.
2. let $\mathbb{F}_{plus}$ for selected features and $\mathbb{F}_{minus}$ for removed features
3. Initialize $\mathbb{F}_{plus} = \Phi$ and $\mathbb{F}_{minus} = \Phi$
4. Choose different classifiers to evaluate the detection rate and the false positive rate
5. Loop over i
   a. Remove feature $f_i$ from $\mathbb{D}$ and its related values as well
   b. If the detection rate and the false positive rate outside a threshold margin
      Then $\mathbb{F}_{plus} \leftarrow \{f_i\}$
      Else $\mathbb{F}_{minus} \leftarrow \{f_i\}$
   c. i = i+1
6. Repeat until $\mathbb{D}$ is empty
7. $\mathbb{F}_{plus}$ has the selected features that affect positively the detection rate and the false positive rate.
8. $\mathbb{F}_{minus}$ has the features that not affect the detection rate and the false positive rate.

---

The above modified SBS infer features according to a threshold margin, which is initiated from the mean value and the standard deviation of the detection rate and the false positive rate of each mentioned classifiers.
Considering the mean value is μ and the standard deviation is σ then the threshold margin is defined as TM = [μ-σ , μ+σ], then

$$\mu = \frac{1}{N}\sum_{i=1}^{N} x_i \text{ and } \sigma = \sqrt{\frac{\sum_{i=1}^{N}(x_i - \mu)^2}{N-1}} \quad (1)$$

where $x$ is the input vector and $N$ is the number of samples in a vector $x$. extracted features from SBS will be then ranked using IG method. However, information gain method only handles discrete values. Therefore transferring continuous values into discrete values is essential to achieve realistic ranks. In this regard, an equal frequency discretizer [21] is considered. It divides all values between minimum and maximum of the feature into k groups (Bins) with equal occurrences. In this work, the number of bins has been initialized to 20 which is the best value came up in our test, so that instances are divided equally.
Let us suppose that $X$ and $Y$ are discrete random variables, $I(X;Y)$ is the information gain of a given attribute $X$ with respect to the class attribute $Y$. When $Y$ and $X$ are discrete variables that take values in $\{y_1,\ldots,y_k\}$ and $\{x_1,\ldots,x_l\}$ with the probability distribution function $P(X)$, then the entropy of $X$ is given by

$$H(X) = -\sum_{i=1}^{l} P(X = x_i)\log_2(P(X = x_i)) \quad (2)$$

Hence, the information gain of feature F on the dataset D in our proposed method is defined as

$$IG(D,F) = H(D) - \sum_{Attr \in Value(F)} \left[ \frac{|D_{Attr}|}{|D|} * H(D_{Attr}) \right] \quad (3)$$

where $Value(F)$ is the set of possible value of $F$, $D_{Attr}$ is the subset of $D$ where $F$ has value $Attr$, $H(D)$ = entropy of class attribute and |.| donates cardinality.

## 3.2. Hybrid Normalization

As mentioned before the network data have several features. These features can be numeric or nominal values [11], [12], those together form a dataset, which will be utilized to train the classifier and to detect abnormal traffic. But most classifiers, particularly Self Organizing Map (SOM), handle only numeric data type and operate properly with the numerical values [17]. Moreover, nominal features such as *protocol type* or *service type* are very important for classification methods especially by neural networks. These classifiers build a model only from numerical data type. Therefore, if symbolic values are not transferred into real values, then the classifier ignores them. Consequently, this will affect the performance of the classification, make the network more vulnerable, and lead perhaps to increase the abnormality. In this paper, the second proposed enhancement scales the nominal values into numeric one using the fact that a symbolic feature describing network traffic can be considered as a sampling of a discrete random variable.

The network traffic can be formally described with $M$ feature vectors $x=(x_1,\ldots,x_n) \in \Omega_1 \text{x} \ldots \text{x} \Omega_n$ of finite dimension $n \in N$, where each element represents a specification of a discrete random variable $X_j$ with values from the probability space $\Omega_j$, $j= 1,\ldots,n$. Let $X_j$ be a random variable with nominal values and $x_{1j}, x_{2j},\ldots,x_{Mj}$ be samples of scale $M$ with the $K$ nominal appearances $nom_{1j}, nom_{2j},\ldots,nom_{Mj}$. Furthermore let $r_{kj}\in N$ be the frequency of occurrences of $nom_{kj}$ into the sampling, then for each $k=1,\ldots,K$ we get

$$r_{kj} = \left|\left\{ i \in N \mid x_{ij} = nom_{kj}, i = 1,..,M \right\}\right| \quad (4)$$

The value $r_{kj}$ is also called absolute frequency of occurrences $nom_{kj}$ into the sampling of scale $M$. We have

$$\sum_{k=1}^{K} r_{kj} = M, 0 \le \frac{r_{kj}}{M} \le 1, k = 1,...,K. \quad (5)$$

hence the values $f_{kj} := \frac{r_{kj}}{M}, k = 1,\ldots,K$ represent the relative frequency of occurrences $nom_{kj}$ into the sampling. Using these relative frequencies we define a mapping $pmf:\Omega_j \rightarrow [0,1]$ that transfers each nominal feature $x_{kj}\in \Omega_j$ into a real number $(x_{kj}) = f_{kj}$.

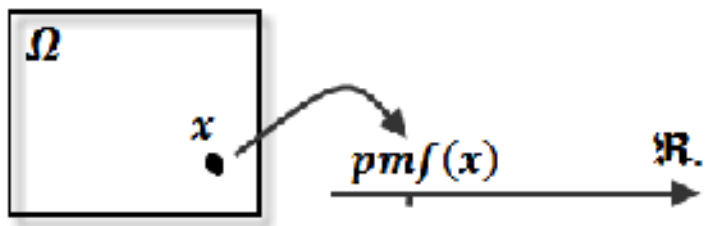

Figure 2. Mapping function of random variable

The proposed enhancement treats only the nominal features. Therefore, the aggregated traffic will be first prepared so that the nominal features are isolated from the numeric one. Figure 1 shows the proposed methodology for nominal features.

For example let $X_2$ be a nominal feature which is *'protocol_type'* and it has the following string values {TCP,UDP,UDP,UDP,RTP,RTP,ICMP,TCP,TCP}, then we have "M=9 and K=4" different strings in the feature. By using the proposed function *pmf* we can calculate the relative frequency that matches each string in the real space. So, *pmf(TCP)=3/9=0.33, pmf(UDP)=3/9=0.33, pmf(RTP)= 2/9=0.22, pmf(ICMP)=1/9= 0.11*. As a result the nominal feature have been transferred to the following numeric representation
*pmf*($X_2$)={0.33,0.33,0.33,0.33,0.22,0.22,0.11, 0.33,0.33}.

Based on this enhancement we could scale the nominal features into real ones into the range [0,1], which is normalized in nature. Thus, it solves the drawback of different feature types and present a feature set with only numeric values. According to this enhancement, an online solution to map the nominal features into numeric in real time could be a sliding window that aggregates the traffic, generates small datasets e.g. 5 seconds length, converts the nominal values into numeric ones directly, and then push the dataset into a queue for further processing.

As mentioned above, most classifiers in IDS need a numeric dataset to detect anomaly traffic sufficiently. However, they suffer from the different scales between features. Feature scale which has large numeric value will be the dominant of any process and small values of other features devolved to be ineffective. Therefore scaling all features into one scale such as between [0, 1] is a necessary step to handle each feature exactly like others. Generally, every dataset consists of nominal and numeric features, the second enhancement handles only the nominal features and scale them directly into the range [0,1], but for the numeric features we examine here three widely used normalization methods, which, in turn, normalize the numeric features. As a result, we present a hybrid approach, which is a modification on the second enhancement, i.e. the proposed transferring method and the normalization of numeric values. Let $f:\mathfrak{R}\rightarrow[0,1]$ be the normalization function and $v\in\mathfrak{R}$ the numerical value of a feature in the feature sets. We denote *nv* to be the normalized feature value after normalization process. Then we can consider the following widely used normalization methods in Table 1

Table 1. Normalization Methods

| **Normalization Method** | **Formula** |
|---|---|
| Decimal Normalization. | $nv = f_1(v) = \frac{v}{10^e}$ <br> where *e* is the minimum number of positions such that maximum value drop into [0,1]. |
| Minimum Maximum Normalization. | $nv = f_2(v) = \frac{v-\min(v)}{\max(v)-\min(v)}$ <br> where min(*v*) and max(*v*) are the minimum and maximum values of feature *v* |
| Statistical Normalization. | $nv = f_3(v) = \frac{v-\mu}{\sigma}$ <br> where μ and σ are the mean value and standard deviation of a feature vector respectively. |

In summary, the modified enhancement will parse the aggregated network traffic and separates the nominal features from the numeric features. Nominal features will be transferred to numeric by the second enhancement, on the other hand, numeric features will be normalized by the best of normalization methods in Table 1. Finally, the seperated nominal and numeric features will be joined again to form a normaloized numeric dataset. See Figure 1.

## 4. RESULTS AND DISCUSSION

The second proposed enhancement in this paper depends on the first enhancement, i.e. results of the first enhancement will be used in the second enhancement. Hence, in this section an illustration about results for each enhancement is presented separately. Moreover, we clarify the steps of preparing and preprocessing the training and testing datasets for the evaluation step. Test datasets are completely separated from the training datasets. Consequently, we have prepared these training and testing datasets with only two class labels; that is, normal and anomaly.

### 4.1. Significant Feature Sets

To evaluate the first enhancement, an offline dataset with 41 features is considered here. In this research work, the KDD cup 1999 is considered, which includes several types of attacks and normal traffic as well. This dataset has become widely a reference dataset for researchers in the area of network security so they can evaluate their models and methodologies. However, this dataset is old and need to be updated. In this regard, [2] proposed an online streaming feature selection that presents a novel framework based on feature relevance. But, the proposed enhancements in this paper will be evaluated with the offline refined dataset NSL-KDD, which is a modified dataset from KDD cup 1999 and it has 41 features and a class that labels each instance as normal or anomaly. Firstly, the dataset with all features have been evaluated by the modified SBS method, and features that fulfill the threshold margin condition are selected. Figure 3 shows a sample result for the detection rate of the classifier J48 by using the modified SBS method.

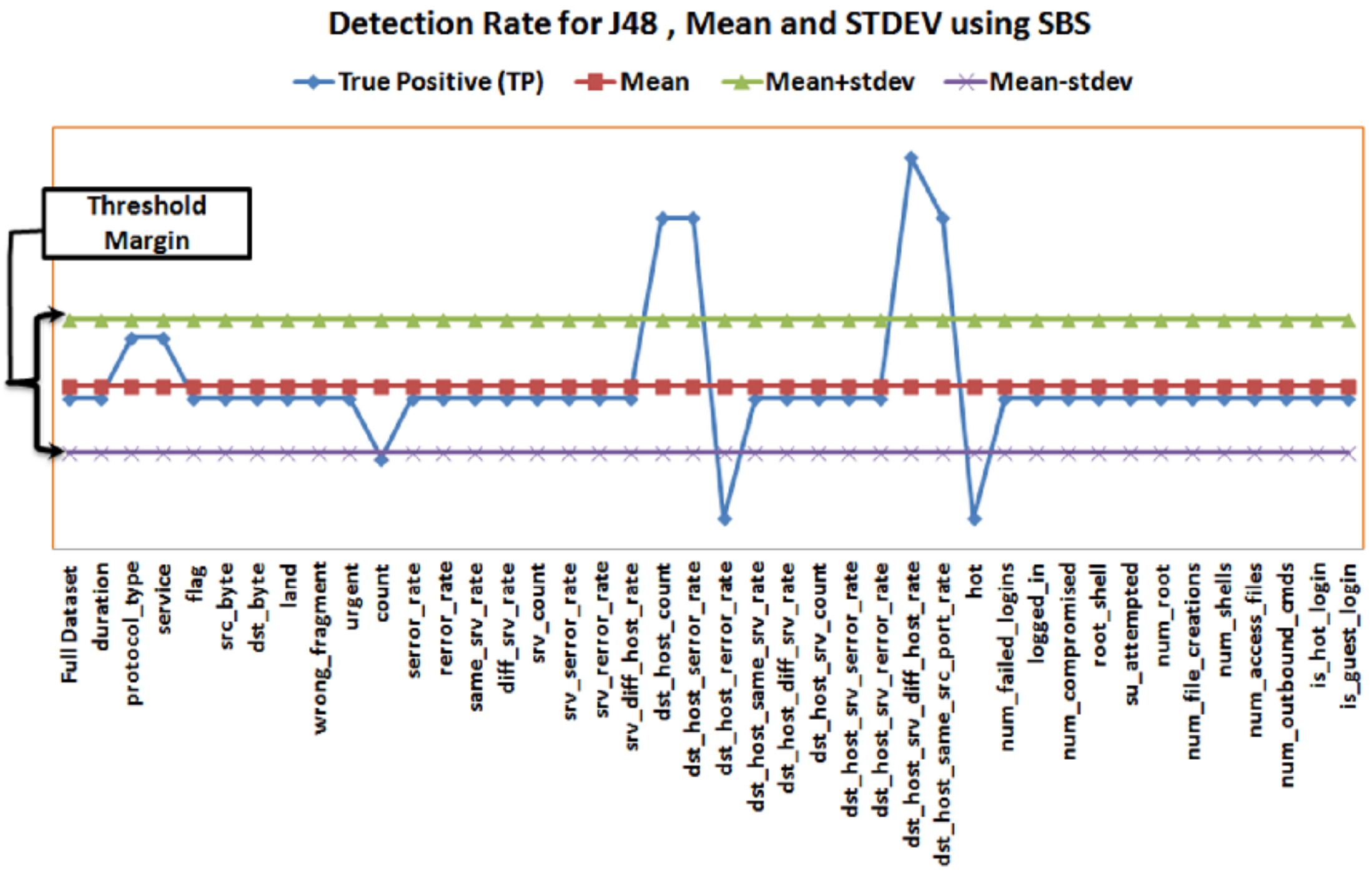

Figure 3. Detection rate result for J48 using SBS method

The middle line on the figure shows the mean value, line with tetragon shape is the detection rate value. For example removing the feature *hot* from the dataset affects the J48 detection rate negatively, it decreases the detection rate and drags it out of the threshold margin. This implies that, this feature is valuable. So, for the classifier J48 there are almost 3 features (see figure 3) valuable for detection rate. Thus, from all classifiers in this work, there are features affecting the detection rate and other the false positive rate, we combine them together to elicit one common feature set.

As a result, the modified SBS in this work is eminent than other methods because it considers both the detection rate and the false positive rate as an evaluation metrics. From SBS, two common feature sets have been selected so that these feature sets improve the detection rate and decrees the false positive rate. However, further improvement to these feature sets is applied, that is, the IG method. This method will calculate the uncertainty of a feature with respect to the class value, i.e. the large IG value of a feature, the most valuable information it delivers.

Thus, extracted features from SBS are first discretized using Equal Frequency Discretization to prepare only discrete dataset and then the IG value is calculated using WEKA Tool 3.7. [18]. Threshold value in IG method is set manually from an expertise to $10^{-4}$. Significantly, we consider two feature sets that deliver valuable information about the network and affect positively the classification metrics. These feature sets are called, Most Valuable Features (MVF) and Most Valuable and Relevant Features (MVRF).

The MVF feature set contains all features fall outside the threshold margin for all classifiers. In contrast, MVRF feature set contains all features fall outside or on the threshold margin for all classifiers. Table 2 shows all features in each feature set.

Table 2. MVF and MVRF sets

| **Name of feature set** | **features** |
|---|---|
| Most Valuable Features (MVF) | service, src_bytes, dst_host_serror_rate, serror_rate, dst_host_srv_diff_host_rate, Protocol_type, rerror_rate, srv_rerror_rate, wrong_fragment, num_compromised, num_access_files |
| Most Valuable and Relevant Features (MVRF) | service, src_bytes, diff_srv_rate, same_srv_rate, dst_host_srv_count, logged_in, dst_host_serror_rate, serror_rate, srv_serror_rate, dst_host_srv_diff_host_rate, protocol_type, rerror_rate, srv_rerror_rate, hot, wrong_fragment, num_compromised, num_access_files, root_shell, num_failed_logins |

The first enhancement using the modified SBS and the IG infer the most valuable features in the intrusion detection system area, such that the detection rate is improved and the false positive rate is mitigated. However, as mentioned before, in unsupervised learning some classifiers handle only features with the same scale, i.e. features values must be between [0,1]. Accordingly, the datasets with only MVF or MVRF should be normalized.

## 4.2. Dataset Types for Evaluation

In the last subsections, we have mentioned precisely which features will be selected and which normalization methods will be considered in this paper. Hence, in this subsection we state different datasets, which are chosen then to evaluate the hybrid approach and to initiate an intensive and comparative study. Thus, we developed a Matlab program “GENDA” to generate several datasets from the MVF and MVRF (see Table 2), normalize them using the hybrid normalization, which maps the nominal features into numeric ones, and the other three normalization methods. That means, GENDA will generate 4 training datasets and 4 testing datasets normalized by the above mentioned normalization methods for the MVF and the same for the MVRF. Figure 4 shows datasets generation using GENDA.

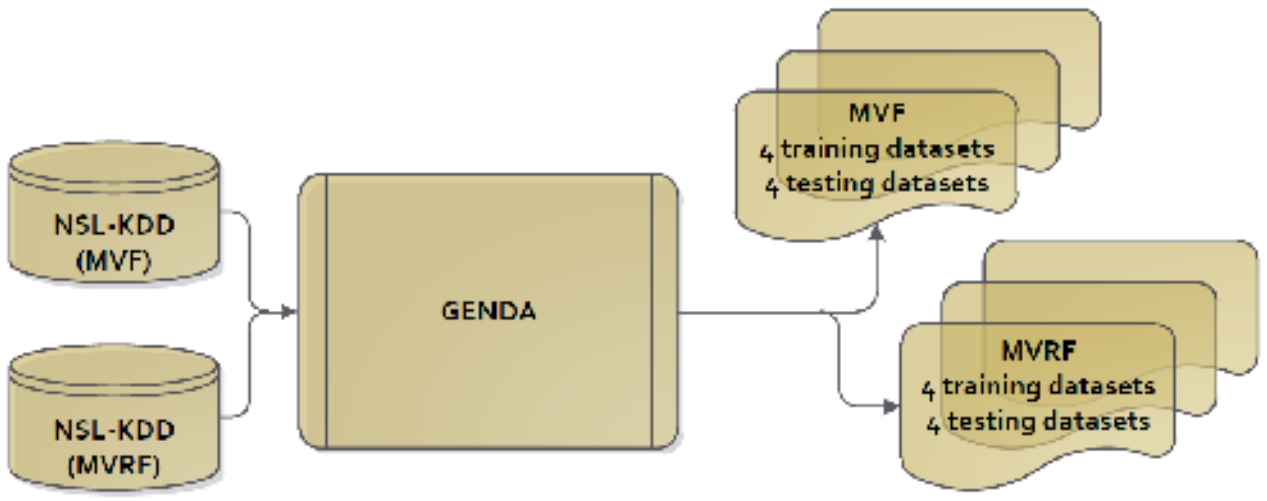

Figure 4. Datasets generation using Matlab

Datasets names and abbreviations are listed in table 3, where L stands for learning dataset (training dataset), T for testing dataset, PMF is the abbreviation for the proposed enhancement for nominal features, N for normalization, the minus sign means "without", plus sign means "with", D for decimal normalization, M for minimum maximum normalization, and S for statistical normalization. For example, L_MVF+PMF+DN means, this is a training dataset, created from the MVF, where the nominal features are normalized by the second proposed enhancement and the numeric features are normalized by decimal normalization method. In addition to the generated datasets in table 3 we have generated another dataset based on the assumption in [3] to compare it to our results.

Table 3. Generated Datasets For Evaluation

| **Dataset** | **Abbreviation** |
|---|---|
| Most valuable features without normalizations - Learning Dataset, Test Dataset | L_MVF-PMF-N<br>T_MVF-PMF-N |
| Most valuable and Relevant features without normalization - Learning Dataset, Test Dataset | L_MVRF-PMF-N<br>T_MVRF-PMF-N |
| Most valuable features with proposed methodology- Learning Dataset, Test Dataset | L_MVF+PMF-N<br>T_MVF+PMF-N |
| Most valuable and Relevant features with proposed methodology- Learning Dataset, Test Dataset | L_MVRF+PMF-N<br>T_MVRF+PMF-N |
| Most valuable features with proposed methodology and Decimal normalization - Learning Dataset, Test Dataset | L_MVF+PMF+DN<br>T_MVF+PMF+DN |
| Most valuable and Relevant features with proposed methodology and Decimal normalization - Learning Dataset, Test Dataset | L_MVRF+PMF+DN<br>T_MVRF+PMF+DN |
| Most valuable features with proposed methodology and Minimum Maximum normalization- Learning Dataset, Test Dataset | L_MVF+PMF+MN<br>T_MVF+PMF+MN |
| Most valuable and Relevant features with proposed methodology and Minimum Maximum normalization - Learning Dataset, Test Dataset | L_MVRF+PMF+MN<br>T_MVRF+PMF+MN |
| Most valuable features with proposed methodology and statistics normalization - Learning Dataset, Test Dataset | L_MVF+PMF+SN<br>T_MVF+PMF+SN |
| Most valuable and Relevant features with proposed methodology and statistic normalization - Learning Dataset, Test Dataset | L_MVRF+PMF+SN<br>T_MVRF+PMF+SN |

The aim of preparing these datasets is to approve that the proposed normalization method solves the drawback of various feature types, on the other hand, to ensure that normalizing the traffic solves the drawback of data dominancy, and finally to compare accurately the impact of each normalization method on the performance parameters of the classification, which are the detection rate, the false positive rate, and testing time. Therefore, the comparison will take place for each normalized dataset and even for the datasets without any normalization to be sure normalization is an efficient method for classification and affects in fact the performance metrics.

## 4.3. Comparative and Evaluative Study

Before showing the results we present here briefly the test environment and the required components to achieve these results. The test environment includes Intel Dual Core processor 3.2 GHz, 4 GB RAM, Linux Ubuntu desktop and Windows 7 Professional 64-bit, Matlab R 2011a, Weka 3.7., and Datasets mentioned in the previous section all has ARFF format. Learn dataset has 125900 instances and test dataset has 22100 instances.

The results of the performance parameters have been evaluated in WEKA tool [18], which is a data mining tool that consists of several implemented classifiers from different fields. We have selected the following classifiers such that supervised and unsupervised methods are covered in the test, these are Self Organizing Map (SOM) with learning rate 0.5 and 500 iteration, Support Vector Machine (LibSVM) with default parameters in WEKA, Decision Trees (J48) with default parameters, Naïve Bayes (NB), Boosting in Adaboost with 5 classifiers for SOM and 5 classifiers for J48 and with mostly 5 iteration and bagsize of 60, and Bagging with also 5 classifiers for SOM and 5 for J48 with the same iteration and bagsize.

For all test cases we have trained the classifier first then evaluated the model with a test dataset and recorded the detection rate, the false positive rate, and testing time accordingly. For all figures we used the same abbreviation in the X-axis, however for Y-axis there are the detection rate, the false positive rate, and testing time respectively. Figure 5 illustrates the results of the detection rate using SOM for six different datasets (see Table 3). We will use figure 5 as an example to explain and clarify the results, notice that line with stars is the MVRF and line with square is the MVF:

SOM1: detection rate results of self organizing map for the non normalized dataset (MVF-PMF-N, MVRF-PMF-N).

SOM2: detection rate results of self organizing map for the dataset with only the second enhancement (MVF+PMF-N, MVRF+PMF-N).

SOM3: detection rate results of self organizing map for the normalized dataset by the hybrid approach, which is the second enhancement and decimal normalization, (MVF+PMF+DN, MVRF+PMF+DN).

SOM4: detection rate results of self organizing map for the normalized dataset by the hybrid approach, which is the second enhancement and statistical normalization (MVF+PMF+SN, MVRF+PMF+SN).

SOM5: detection rate results of self organizing map for the normalized dataset by the hybrid approach, which is the second enhancement and min max normalization (MVF+PMF+MN, MVRF+PMF+MN).

SOM6: detection rate results of self organizing map for the normalized dataset by the normalization method proposed in [3].

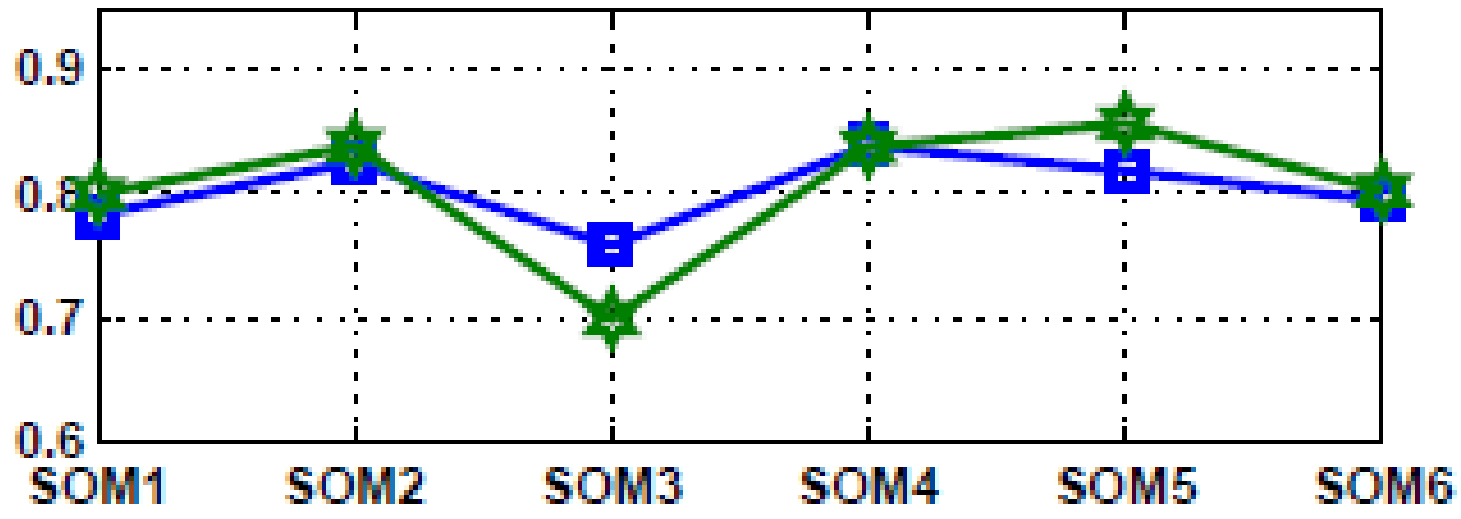

Figure 5. Detection rate result using SOM

The above figure shows the results for the performance parameter detection rate (true positive TP). It shows that SOM5 has the maximum detection rate, which means the dataset with the features of MVRF normalized with the hybrid approach (the second enhancement and the minimum maximum method) using SOM achieves the best detection rate. On the other hand, SOM3 shows the lowest detection rate.

Hence, at the end of this work all classifiers' results regard the three performance parameters for different datasets are illustrated (Figure 6, 7, and 8).

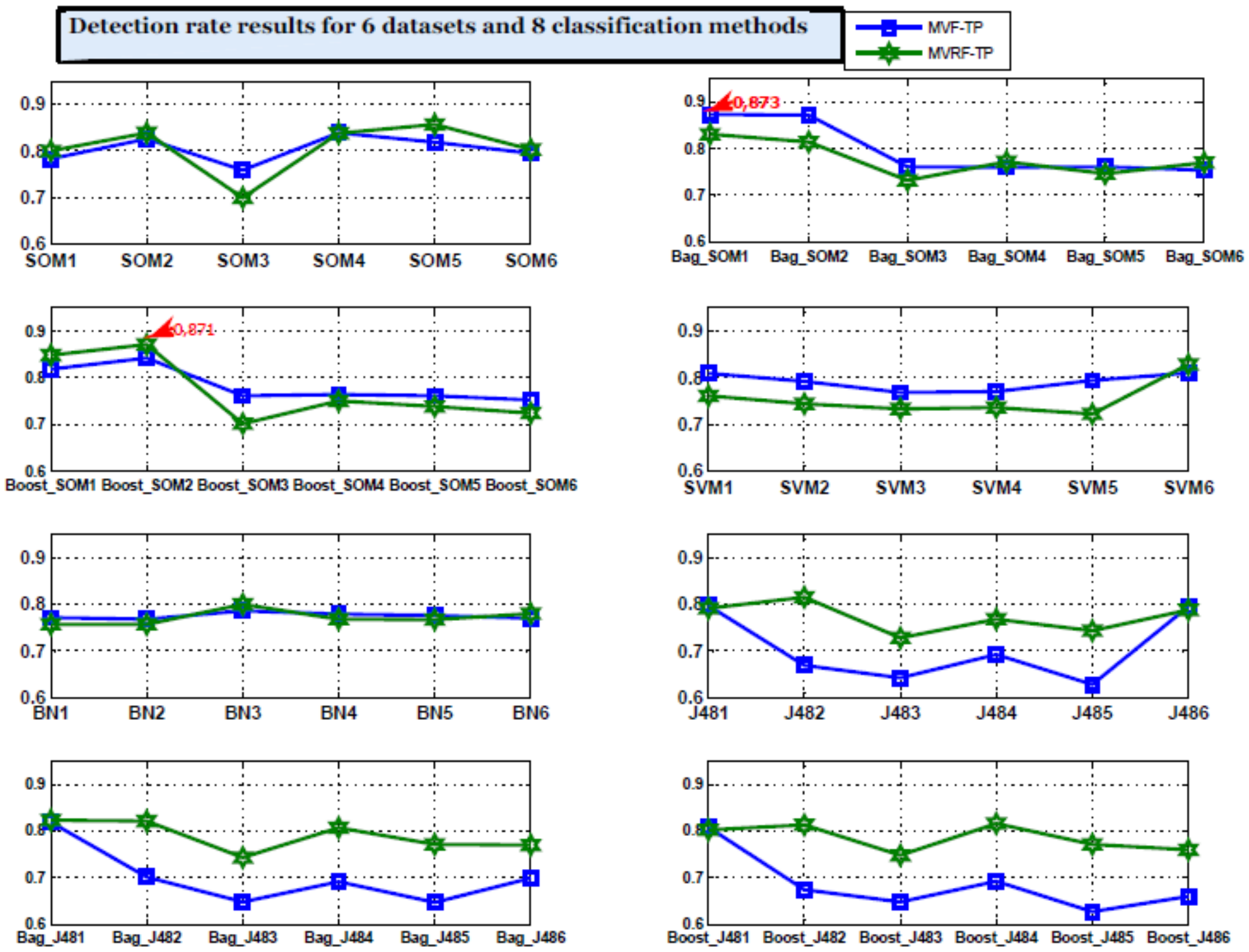

Figure 6. Detection rate result for all classifiers

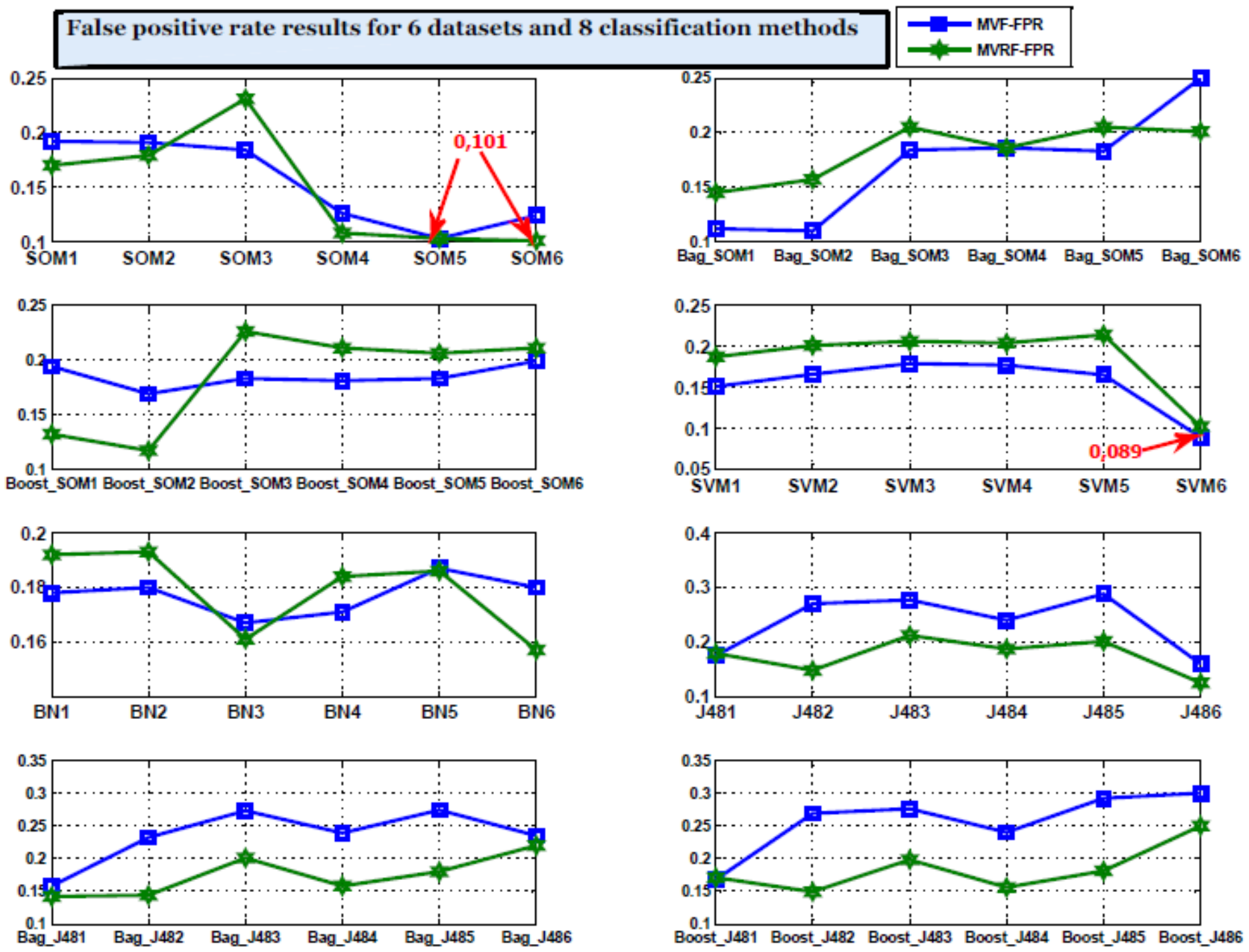

Figure 7. False positive rate for all classifiers

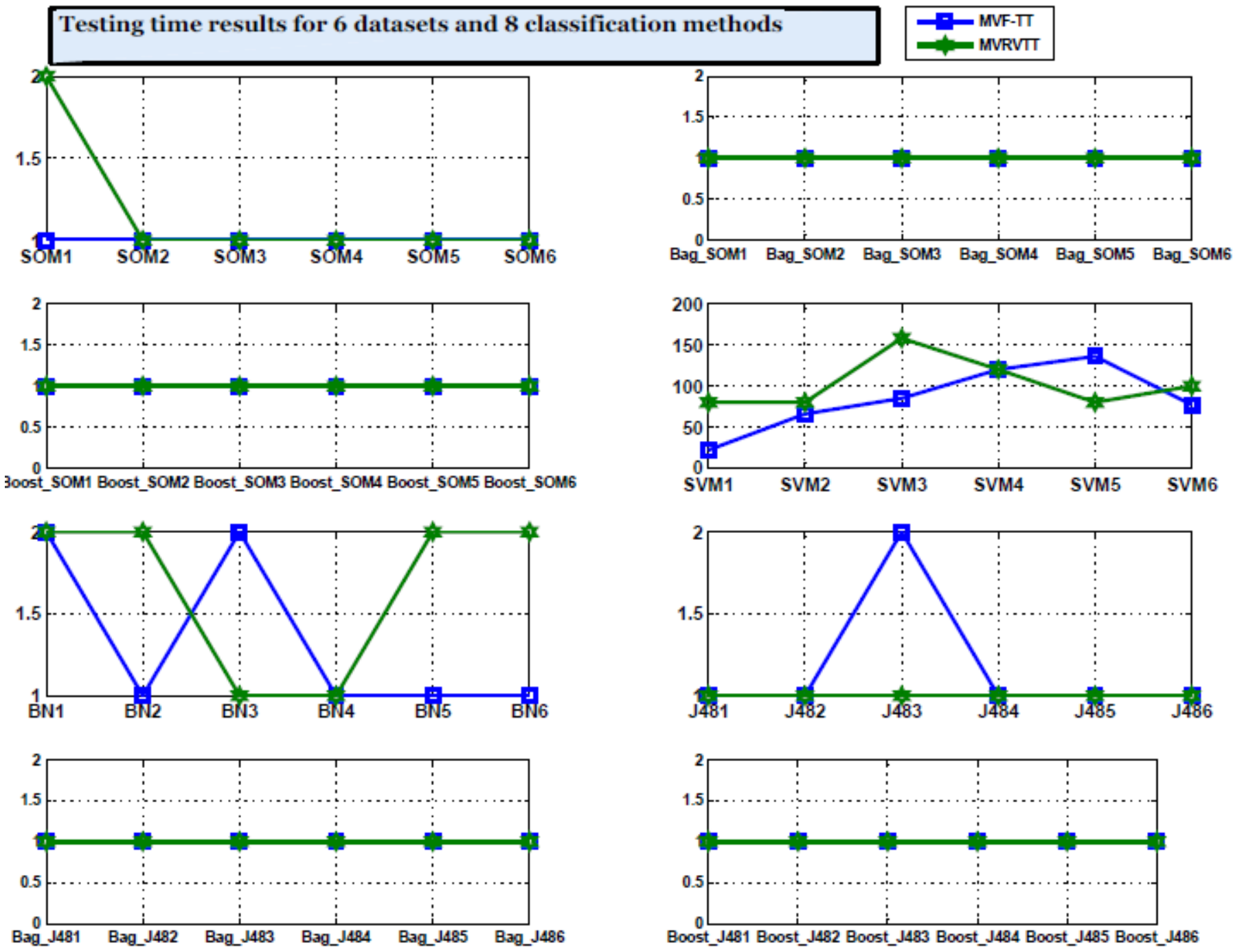

Figure 8. Testing time result for all classifiers

According to these results, which are deduced from the implemented classifiers in WEKA, we summarize the outcome into two main tables which are table 4 and 5, that is, these tables show the final result of the detection rate and the false positive rate for the two feature sets, the most valuable feature set and the most valuable and relevant feature set.

Table 4. Final Results For Detection Rate

| Classifier | MVF | Classifier | MVRF |
|---|---|---|---|
| Bag_SOM1 | 0,873 | Boost_SOM2 | 0,871 |
| Bag_SOM2 | 0,872 | SOM5 | 0,8570 |
| Boost_SOM2 | 0,842 | Boost_SOM1 | 0,848 |
| SOM4 | 0,8390 | SOM2 | 0,8390 |
| SOM2 | 0,8260 | SOM4 | 0,8390 |

Table 5. Final Results For False Positive Rate

| Classifier | MVF | Classifier | MVRF |
|---|---|---|---|
| SVM6 | 0,0890 | SOM6 | 0,1010 |
| SOM5 | 0,1030 | SVM6 | 0,1020 |
| Bag_SOM2 | 0,11 | SOM5 | 0,1030 |
| Bag_SOM1 | 0,112 | SOM4 | 0,1080 |
| SOM6 | 0,1240 | Boost_SOM2 | 0,117 |

The third performance parameter is the testing time. We have not shown here the results of testing time because all classifiers needed only one to two second to evaluate the test dataset on the training model except SVM which needed at least 80 seconds to evaluate the test dataset (see fig. 8).

Obviously, the result confirms that the hybrid normalization approach boosts neural networks to have the best performance parameters specially by using SOM. It is noticeable that the SVM has the best false positive rate but in contrast to the testing time it has the longest one. Therefore, SOM has almost the best performance in all test cases. On the other hand, both datasets, MVF and MVRF, have approximately the same results so we can exploit one of them in IDS. Tables 4 and 5 articulate the best detection rate and the lowest false positive rate, which confirms that, the best normalization method is to normalize the nominal features using the second enhancement (the mapping function *pmf* ) together with statistical or minimum maximum normalization. That is to imply, the hybrid approach achieves the target of eliminating classifications drawbacks and consequently improves the accuracy of classification in neural networks. In contrast to the proposed normalization method in [3] the hybrid approach achieved more accuracy and preciseness with low false positive and smallest testing time.

## 5. CONCLUSION

The huge dimension of network traffic and the variety of feature types degrade the performance of IDS. Moreover, different scales of feature values affect the performance negatively. Thus two significant enhancements are proposed in this paper to eliminate these drawbacks. The first enhancement is an improved feature selection method that includes a modified sequential backward search and a ranking method using information gain. The second enhancements, is a hybrid approach, which consists of transferring nominal values of network features to numeric ones by exploiting a concept that related to the idea of discrete random variable and probability mass function, and then combine this enhancement to a known normalization method such as decimal, statistical, or minimum maximum normalization. Several tools such as WEKA and Matlab are utilized to develop special programs or to evaluate the proposed enhancements. However, in WEKA several classifiers are chosen in the evaluation, which are supervised and unsupervised. The results show that, the first enhancement has improved the detection rate and mitigate the false positive rate in the IDS, which means the feature sets, MVF and MVRF, enhance the performance significantly. On the other hand, the hybrid approach has achieved the best detection rate and the lowest false positive rate in our intensive study. This hybrid approach is consisting of the second enhancement and the statistical or minimum maximum normalization method. In contrast to other normalization approach, we achieve a better performance and results. In the coming research activity, we will concentrate on abstracting several associations between these features to define a normal behavior for the IDS.

## ACKNOWLEDGMENTS

This research project “SecMonet” is supported by the German Federal Ministry of Education and Research (Funding Code 17062X10) under the funding line “ProfUnt”.

## Authors

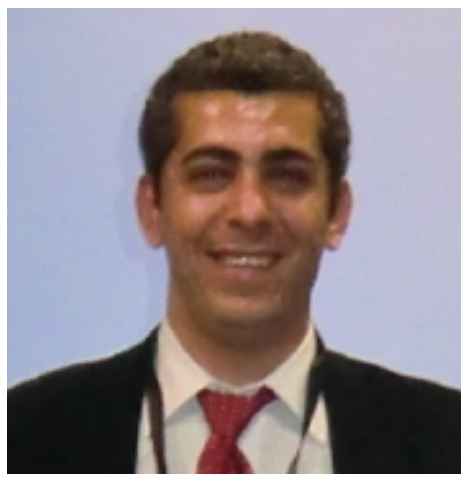

**Maher Salem** is currently a research assistant in the Network and Data Security group at the University of Applied Sciences Fulda in Germany; he is pursuing his Ph.D. in Kassel University in Germany. He received his M.Sc. in Computer Engineering from the University Duisburg-Essen in 2006 (Germany), and B.Sc. in Electrical and Computer Engineering from Jordan University of Science and Technology in 2002 (Jordan). He has more than 10 years experience. He was a system engineer in Nokia networking R&D, System Engineer in E-plus Group for Mobile Networks, and IT-Specialist with other pioneer firms. Additionally, he published several topics in the area of network security, participated in international workshops and exhibitions, and contributed with national and international projects. He is supervising several Master and Bachelor theses which focus on preventing the newly unknown attacks and combating cybercrimes. His main research interests are network security innovations, data mining and machine learning, and network optimization.

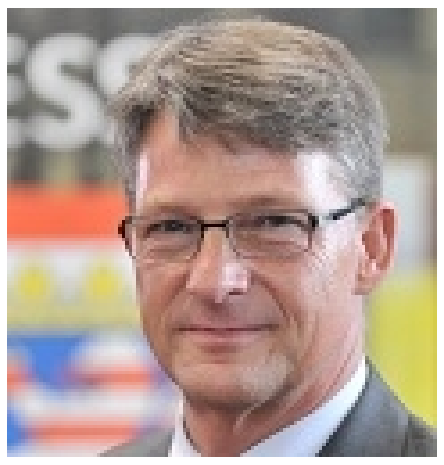

**Ulrich Buehler** is a professor at the University of Applied Sciences Fulda, Germany. Currently, he is the leader of the group of network and data security (NDSec) and supervising several projects such as “SecMonet”. Prof. Buehler worked as a guest professor at several international universities, e.g. UK, USA, and Spain. Since 2004 he is occupying several positions, such as federal chairman of the Board of “German Association of Faculties of Informatics at Universities of Applied Sciences” (FBTI), member of the Board of European Quality Assurance Network for Informatics Education (EQANIE), member of the Executive Board of Accreditation Agency for Study Programs in Engineering, Informatics, Natural Sciences and Mathematics (ASIIN), and other valuable memberships. His research area focuses on network security, cryptography, applied mathematics, and data mining. Moreover, main focuses of his lectures are network and data security, cryptography and applications, formal methods in IT-Security, applied mathematics, and algebra.